\RequirePackage{fix-cm}
\documentclass[twocolumn,epjc3]{svjour3}  
\smartqed  
\RequirePackage{graphicx}
\usepackage{hyperref}       
\usepackage{url}            
\usepackage[LGRgreek]{mathastext}
\usepackage{graphicx}
\graphicspath{{figs/}}
\usepackage{subfig}
\usepackage{multirow}
\usepackage{cite}
\newcommand{\geant}{{\sc Geant4 }}
\journalname{Eur. Phys. J. C}
\begin{document}

\title{Optimising longitudinal and lateral calorimeter granularity for software compensation in hadronic showers using deep neural networks}
\subtitle{}
\titlerunning{Optimising calorimeter granularity for software compensation}        

\author{Coralie Neub\"user\thanksref{e1,addr1} \and Jan Kieseler\thanksref{e2,addr2} \and Paul Lujan\thanksref{e3,addr3}} 


\thankstext{e1}{e-mail: coralie.neubuser@cern.ch}
\thankstext{e2}{e-mail: jan.kieseler@cern.ch}
\thankstext{e3}{e-mail: paul.lujan@cern.ch}


\institute{INFN TIFPA, Via Sommarive, 14 I-38123 Trento, Italy \label{addr1}
           \and
           CERN, 1211 Gen\`eve 23, Switzerland \label{addr2}
           \and
           University of Canterbury, Private Bag 4800, Christchurch 8140, New Zealand \label{addr3}
}

\date{Received: date / Accepted: date}

\maketitle

\begin{abstract}
We investigate the effect of longitudinal and transverse calorimeter segmentation on event-by-event software compensation for hadronic showers. To factorize out sampling and electronics effects, events are simulated in which a single charged pion is shot at a homogenous lead glass calorimeter, split into longitudinal and transverse segments of varying size. As an approximation of an optimal reconstruction, a neural network-based energy regression is trained. The architecture is based on blocks of convolutional kernels customized for shower energy regression using local energy densities; biases at the edges of the training dataset are mitigated using a histogram technique. With this approximation, we find that a longitudinal and transverse segment size less than or equal to 0.5 and 1.3 nuclear interaction lengths, respectively, is necessary to achieve an optimal energy measurement. In addition, an intrinsic energy resolution of $8\%/\sqrt{E}$ for pion showers is observed.
\end{abstract}

\maketitle

\section{Introduction}
Both existing high-energy physics experiments, such as those at the CERN LHC, and future experiments at future colliders, like the Future Circular Collider (FCC), rely heavily on the performance of hadron calorimeters and their particle flow capabilities for measuring jet and missing transverse momentum ($p_{T}$)~\cite{Sirunyan:2017ulk,Aaboud:2017aca,Ruan:2014paa,Thomson:2009rp,Marshall:2012ry,Marshall:2013bda,Marshall:2015rfa,Sefkow:2015hna,Tran:2017tgr}. Hadron calorimeters are currently characterized not only in terms of their intrinsic energy resolution, but by their imaging capabilities, which allow for offline corrections using smart algorithms. Due to the diverse composition of hadronic showers and the differences in the calorimeter response, a correct energy measurement becomes challenging. In general, the components of hadronic showers can be divided into electromagnetic (EM) and hadronic parts. The hadronic part of the shower consists of particles such as neutrinos and neutrons which are partially invisible to the detector. This can be affected by the chosen active detector material, where, e.g., plastic scintillators allow for neutron detection via strong interaction with the atomic nucleus. The undetectable particles in the hadronic shower result in an unequal detector response; that is, $e/h\neq1$, where $e$ and $h$ are the calorimeter response to electromagnetic and hadronic shower fractions, respectively.

Many hadronic calorimeters currently in use and planned for future experiments are sampling calorimeters, which consist of alternating active and passive absorber layers~\cite{CERN-LHCC-97-031,CERN-LHCC-96-041,HGCAL-TDR,Neubuser:2705432}. The sampling of the hadronic shower allows for tuning of the hadronic and electromagnetic shower responses. In the past, the $e/h$ ratio has been adjusted closer to 1 by either suppressing the electromagnetic response, e.g., by using high-$Z$ absorbers, or by enhancing the hadronic response, using neutron-sensitive active materials. Calorimeters that have a ratio $e/h\sim1$ are called ``compensating'' calorimeters. These optimizations in the active and passive materials often require a decreased sampling fraction (ratio of active/passive material), which itself degrades the calorimeter energy resolution by increasing the stochastic term $\alpha$ of  
\begin{equation}
\frac{\sigma_{E}}{\left<E\right>}=\frac{\alpha}{E} \oplus c.
\label{eq:reso}
\end{equation}
The stochastic term is dominated by the sampling fraction and frequency for sampling calorimeters, and expresses the dependence of the calorimeter resolution on the fluctuations of the number of particles within the hadronic shower (following a Poisson distribution). The constant term $c$ expresses energy-dependent uncertainties, like the fluctuations on the EM-to-hadronic shower fraction, which is logarithmically increasing with energy, or energy losses due to particles escaping the detector, caused by limited calorimeter sizes. The first can be removed either by intrinsic compensation, or by an event-by-event measurement of the EM fraction, which is called software compensation. \\
Due to the cost and mechanical stability benefits, absorbers made of steel or lead are widely in use. These materials have been found to require very small sampling fractions in, e.g., scintillator-steel calorimeters in order to achieve compensating behavior. Since such low sampling fractions would degrade the performance, especially for particles at low energies ($<50$\,GeV), the solution to correct for fluctuations in the electromagnetic shower fraction is to use software compensation techniques. \\
In order to allow algorithms to distinguish between the dense electromagnetic shower core and other shower parts, such as e.g. disappearing tracks, the granularity of the calorimeter plays a key role. The first attempt in so-called imaging calorimetry has been made by the CALICE collaboration, which started a R\&D program of calorimeters for a future e$^{-}$e$^{+}$ linear collider~\cite{Israeli:2018byq,Chefdeville:2019zzq}, where the calorimeter designs have been optimised for particle flow algorithms~\cite{Marshall:2012ry}. These algorithms allow for jet energy measurements using the best suited sub-detector to reconstruct each jet sub-particle. The prototypes of these calorimeters have been realised with active layers made of silicon for the EM shower part and scintillator or resistive plate chambers for the measurement of hadronic showers. The active layers were tested and interleaved within both steel and tungsten absorber stacks~\cite{Chefdeville:2015,Adloff:2012gv} and achieved such good results in testbeams~\cite{Eigen:2019ccp} that the CMS collaboration decided to adopt this concept in a full silicon-tungsten/scintillator-steel endcap calorimeter~\cite{HGCAL-TDR,Quast:2017gnq}. The developments in, e.g., silicon photomultiplier (SiPM) technologies have been the key to measure the scintillation light produced in calorimeter cell sizes of $3\times3\times0.5$\,cm$^{3}$~\cite{Sefkow:2018rhp}. The impact of software compensation techniques on the performance of particle flow algorithms has been studied in a specific detector design~\cite{Tran:2017tgr}, and proven to provide a significant improvement to the jet energy measurement by using a corrected calorimeter cluster which is matched to tracks in the tracking system. \\
The next step towards a calorimeter design optimized for the use of software compensation techniques is to study the necessary granularity that allows an algorithm to determine most accurately the hadronic shower energy.  


In this paper, we will discuss the performance of a software compensation technique using a deep neural network (DNN), with a specific focus on the dependence on the transverse and longitudinal granularity. Therefore, a homogenous model calorimeter has been studied in full \geant simulations. The performance is evaluated in terms of energy resolution and linearity for single charged pions. The goal is to determine the minimal granularity of a calorimeter needed to achieve the best energy measurement using DNNs. As the choice of granularity can influence the detector design and cost, a measurement of the impact of this choice is necessary in order to optimize the design. Here, the DNN is utilised as a generic close-to-optimal reconstruction algorithm that can be optimised to the granularity in an automatised fashion.

\section{Calorimeter and dataset}
The studied calorimeter is a homogeneous lead tungstate calorimeter, which follows the EM calorimeter concept of the CMS experiment~\cite{CERN-LHCC-97-033}. However, we do not consider any passive absorber material, assuming that the impact on the calorimeter performance of the sampling fraction and the longitudinal and transverse segmentation are uncorrelated. The dimensions are $1\times1\times2.5$\,m$^3$, which ensures complete shower containment within the calorimeter volume and corresponds to $10\,\lambda$ and $200\,X_{0}$ of total depth.
The longitudinal and transverse segmentation is increased from no segmentation up to $30\times30$ segments in $x$ and $y$, and from 1 to 60 segments in the lateral direction. A list of the configurations can be found in Table~\ref{tab:dataset}.

The data set consists of approximately $5\times10^6$ charged pion events, generated using the FTFP\_BERT physics list of \geant 10.04 patch 0. The training data set comprises pions with energies sampled from a flat distribution between 1 and 110\,GeV. The test data set
covers 11 discrete energies of 5 to 105\,GeV in 10\,GeV steps. The \geant simulation has been performed in the highest granularity, while for the tests and training of different segmentation configurations, the same dataset has been used. For this purpose, the energy deposits in the cells have been merged corresponding to the tested cell sizes. This method avoids inconsistencies that are otherwise to be expected due to the different number of surfaces and material borders through which \geant propagates the particles.

\begin{table}[htp]
\begin{center}
\begin{tabular}{c|c|c|c|c}
\multirow{2}{*}{Stage} &  Longitudinal & \multicolumn{3}{c}{Depth of layers} \\
   &   segments 	& in cm  	&  in ${\lambda}_{\pi}$	& in $X_0$ \\ 
\hline
0	&	1			& 250			& 9.8					& 198			\\
1	&	6 			& 41.7			& 1.6					& 33				\\
2	&	10			& 25				& 1.0					& 20				\\
3	&	12			& 20.8			& 0.8					& 16				\\
4	&	15			& 16.7			& 0.7					& 13				\\
5	&	20			& 12.5			& 0.5					& 10				\\
6	&	30			& 8.3				& 0.3					& 7				\\
7	&	60			& 4.2				& 0.2					& 3				\\
\multicolumn{5}{c}{} \\
\multirow{2}{*}{Stage} & Transverse & \multicolumn{3}{c}{Size of cells} \\
		& segments   & in cm$^2$ 	& in $\lambda_{n}$	&  in $X_{0}$	\\
\hline
default	& $1\times1$ 		& $100\times100$	& $3.9\times3.9$		& $79\times79$		\\  
A		& $3\times3$		& $33\times33$ 	& $1.3\times1.3$ 		& $26\times26$ 	\\
B		& $5\times5$		& $20\times20$ 	& $0.8\times0.8$ 		& $16\times16$ 	\\
C		& $10\times10$		& $10\times10$ 	& $0.4\times0.4$ 		& $8\times8$ 		\\
D		& $15\times15$		& $6.7\times6.7$ 	& $0.3\times0.3$ 		& $5\times5$ 		\\
E		& $30\times30$		& $3.3\times3.3$ 	& $0.1\times0.1$ 		& $3\times3$ 		\\
\end{tabular}
\end{center}
\caption{Granularity configurations considered in this analysis.}
\label{tab:dataset}
\end{table}%

\section{Neural network architecture and training}

At the core of the neural network architecture used here is a software compensation block that uses convolutional neural network (CNN) layers~\cite{lecun1998gradient} to achieve local identification of the subshowers, similar to the one introduced in Ref.~\cite{Neubuser:2705432}, which is used as a subblock in the overall model. This subblock consists of 3 parallel paths: in the first path, the energy of all cells within the kernel range $K$ is summed up and forwarded to the next block, while this kernel is moved with a stride of size $K$; the second path consists of a CNN layer with the same kernel size and $F=16$ filters; the third path contains in total three subsequent CNN layers, out of which the first two have kernel sizes (in x, y, and depth) of $K_a = (1,\ k,\ 3)$ and $K_b=(k,\ 1,\ 3)$, with no stride and 32 filters, each. Here, $k$ is an adjustable parameter depending on granularity, as described later. The final layer of this path is a CNN layer with a kernel size of $K$ with a stride of $K$ and $F$ filters, such that the output of all paths can be combined. This combination is done by adding the output of the CNN layers of all paths feature by feature. All layers use a tanh activation function. The weights of the layers in the third path are initialised with a Gaussian distribution centred at 0 with a width of $10^{-3}$, and receive a small $L2$ regularisation of $10^{-5}$. This structure is optimised to derive small corrections to the simple energy sum by detecting the different shower shape of electromagnetic subshowers.


In the final model, the input is passed through a batch normalisation layer~\cite{ioffe2015batch}, normalising all inputs except for the per-cell energy. If less than 6 calorimeter layers are present or the transverse granularity in either direction is less than 6, the input is directly flattened and passed to 3 dense layers, the first two of which contain 128 and 64 nodes using ELU activation~\cite{elu_activation}, before being finally passed to the energy prediction layer with 1 node. In all other cases, the input is first passed through a set of the subblocks described above before being fed through the same structure with dense layers. 
These subblocks adapt to the input: if the corresponding granularity is less than $6\times6$ cells in the transverse directions, a stride of $1\times 1$ is used, and the input $k$ for the kernel size determination is set to $k=1$. Otherwise, a stride of $2\times 2$ and $k=3$ are used in these directions. The subblock is repeated until the dimensionality in $x$, $y$, or depth is less than or equal to 6. At this point, the output is fed to the three final dense layers.

The model is trained using the Adam optimiser~\cite{kingma2014adam} using TensorFlow~\cite{tensorflow} and Keras~\cite{keras} within the DeepJetCore framework~\cite{DJC}. The training consists of five steps: the first four steps use a loss function $L_\mathrm{calo}$ that follows the expected calorimeter resolution:
\begin{equation}
    L_\mathrm{calo} = \frac{(E_\mathrm{true} - E_\mathrm{pred})^2}{E_\mathrm{true}} \mathrm{.}
\end{equation}
These steps are trained for 1, 19, 60, and 20 epochs with learning rates of $10^{-4}$, $10^{-4}$, $10^{-5}$, and $10^{-5}$, and batch sizes of 256, 512, 1280, and 1280. Between the third and fourth step, the batch normalisation is frozen. 

The mean and expectation value for $E_\mathrm{true}$ differ at the edges of the training sample. This typically leads to edge effects, which introduce a bias towards higher predicted values at the low edge, and towards lower predicted values at the high edge. 

To mitigate the effect, we freeze all layers except for the last dense layers, and introduce a loss that follows a $\chi^2$ distribution taking the difference of the average predicted and truth energy in bins of $E_\mathrm{true}$, and accounting for the number of samples in that bin. The bin boundaries are randomly chosen for each batch to avoid a global bias. Using this loss, the model is trained for another 50 epochs with a learning rate of $10^{-5}$ and a batch size of 1280.

\begin{figure}[htbp]
\begin{center}
\subfloat[\label{fig:stage4_distr}]{\includegraphics[width=0.45\textwidth]{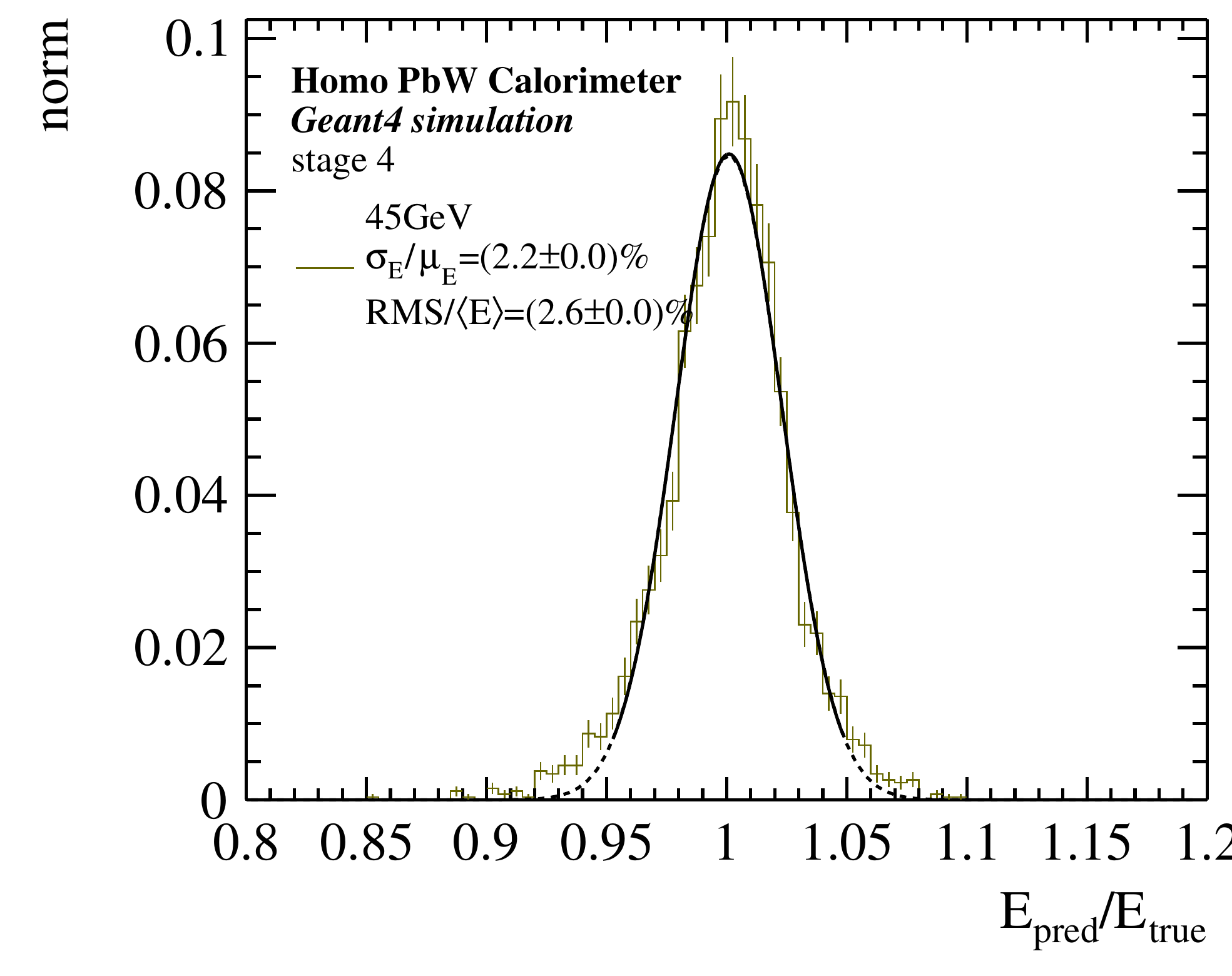}}\\
\subfloat[\label{fig:stage4_reso}]{\includegraphics[width=0.45\textwidth]{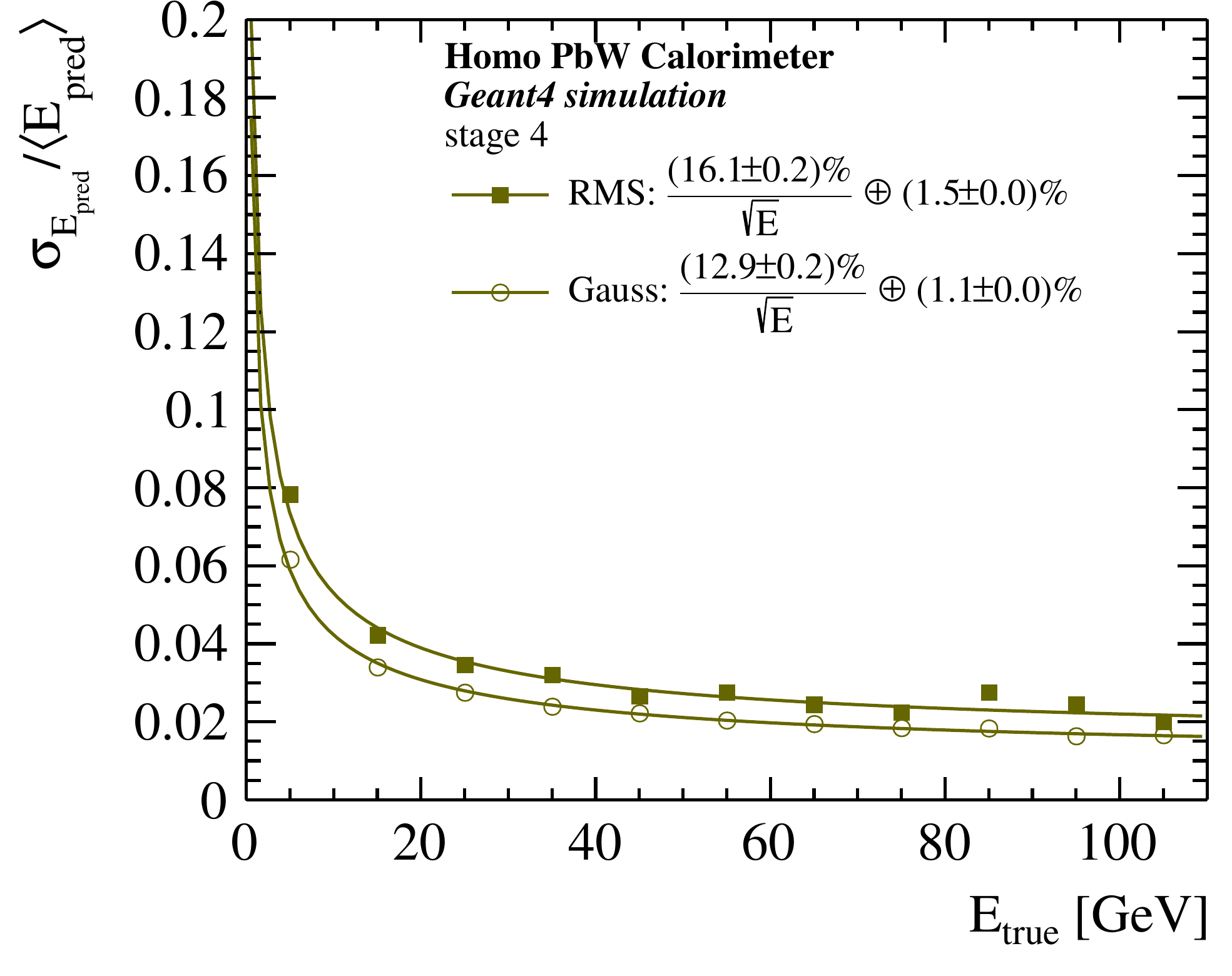}}
\caption{Results for a scenario with 15 longitudinal layers (stage 4) and no transverse segmentation. (a) Energy distribution for 45\,GeV pions. The width as computed from the Gaussian fit (black line) and from the RMS are shown. (b) Energy resolution as a function of the particle energy. The resolution is computed two ways, using the Gaussian fit (open circles) and using the RMS (filled squares).}
\label{fig:stage4}
\end{center}
\end{figure}

\section{Results}
The energy resolution is evaluated as the ratio of the width to the most probable value of the distribution of the reconstructed energy. These distributions, as shown for example in Figure~\ref{fig:stage4_distr}, follow a Gaussian function. The standard deviation can thus be extracted from a fit. This fit is limited within 2$\sigma$ around the most probable value $\mu$, following the procedure widely used in calorimeter performance studies. As a comparison and validation, the energy resolution has also been evaluated from the root mean square (RMS) and mean, which is sensitive to the tails of the distribution. The energy resolution over the full available energy range is shown for stage 4, which corresponds to a granularity of 15\,longitudinal layers, in Figure~\ref{fig:stage4_reso}. The points are fitted following Equation~\ref{eq:reso}, and the values of the stochastic and constant term are shown in the legend. An overall 10--20\% degradation in energy resolution from the Gaussian fit to the RMS method is observed.  \\
In the following, the energy resolutions obtained for different granularities will refer to the results obtained from the Gaussian fit.

The results, in terms of the stochastic term $\alpha$ and constant term $c$ for all studied longitudinal and transverse granularities, are summarized in Table~\ref{tab:results_reso}. The theory of the different contributions to the energy resolution of hadronic showers~\cite{Fabjan:1989ti} considers that the stochastic term is in fact a sum of two major effects, $\alpha = \alpha_{\mathrm{int}}\oplus\alpha_{\mathrm{sampl}}$, where the first intrinsic term is irreducible and determined by the fluctuations of the initial energy that is transformed into ionising shower particles, and the second is the term due to the sampling fraction. These losses are material dependent, due to material-dependent nuclear binding energy losses, and have been found to be on the order of 19\%/11\% in the ZEUS uranium/lead-scintillator calorimeter prototypes~\cite{Tiecke:1989nz}. \\

\begin{figure}[htbp]
\begin{center}
\subfloat[]{
\includegraphics[width=0.45\textwidth]{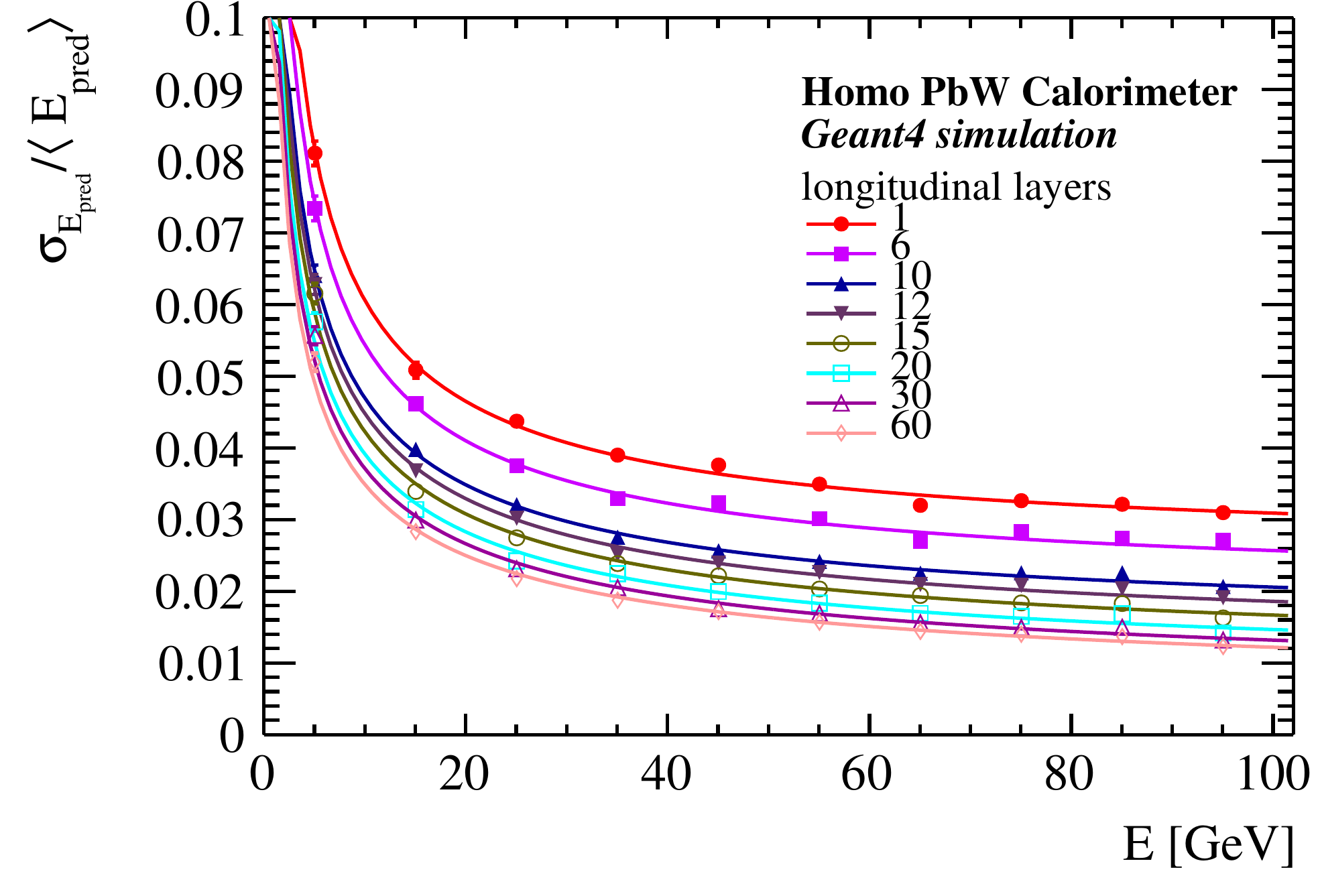}
}\\
\subfloat[]{
\includegraphics[width=0.45\textwidth]{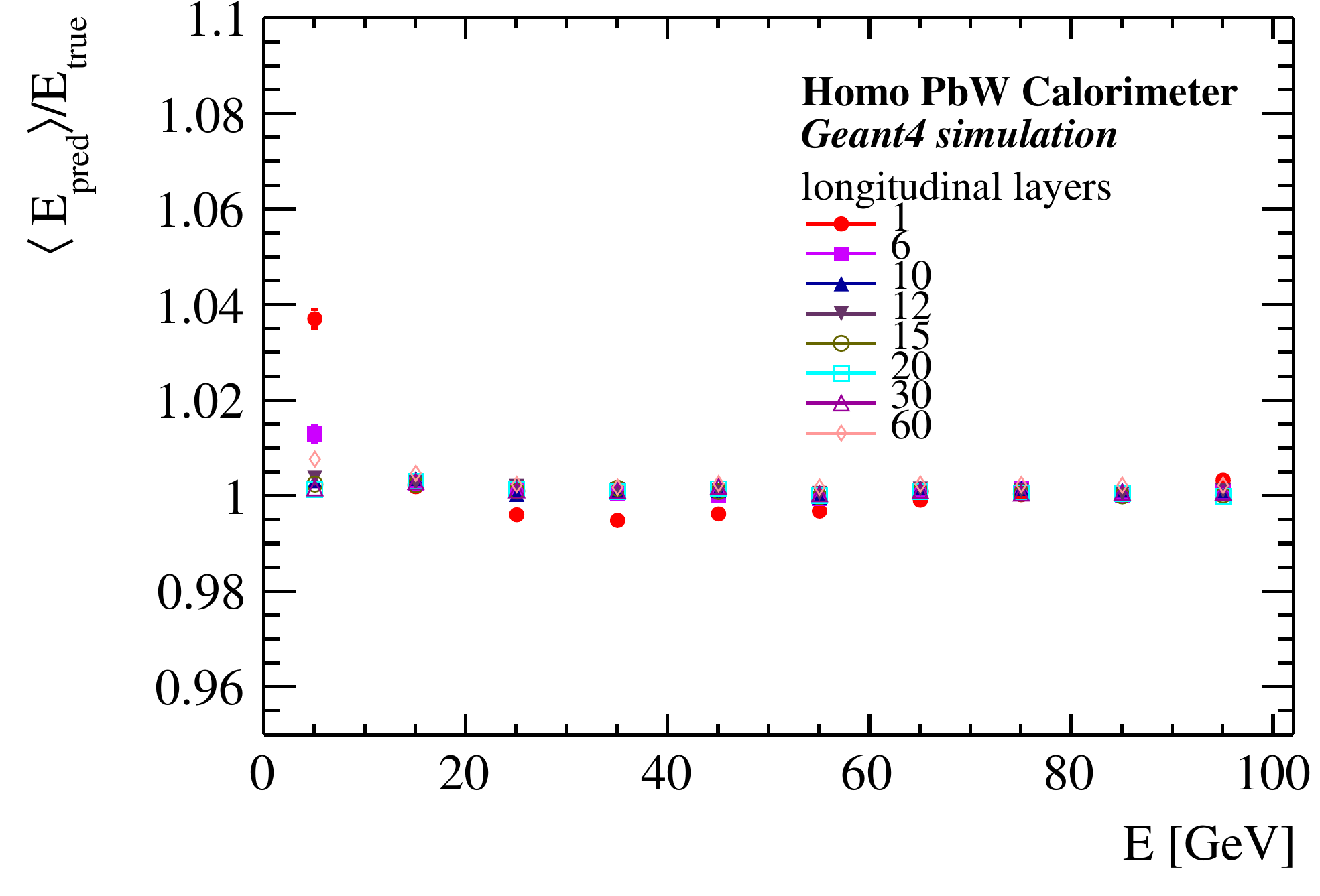}
}
\caption{Energy resolution (a) and linearity (b) for different longitudinal granularities and no transverse segmentation. The curves correspond to the fit with Equation~\ref{eq:reso}.}
\label{fig:res}
\end{center}
\end{figure}

\begin{figure}[htbp]
\begin{center}
\subfloat[]{
\includegraphics[width=0.45\textwidth]{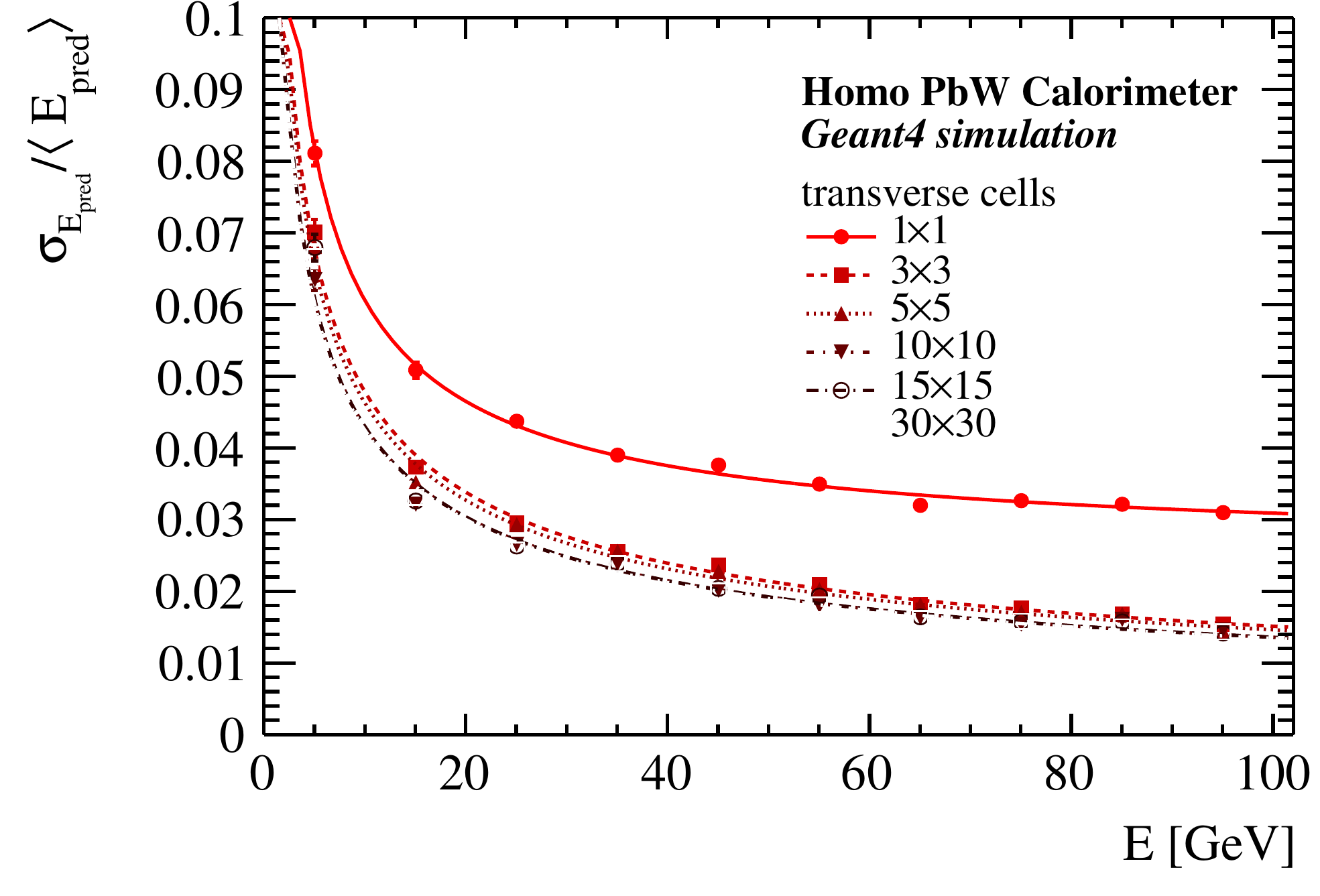}
}\\
\subfloat[]{
\includegraphics[width=0.45\textwidth]{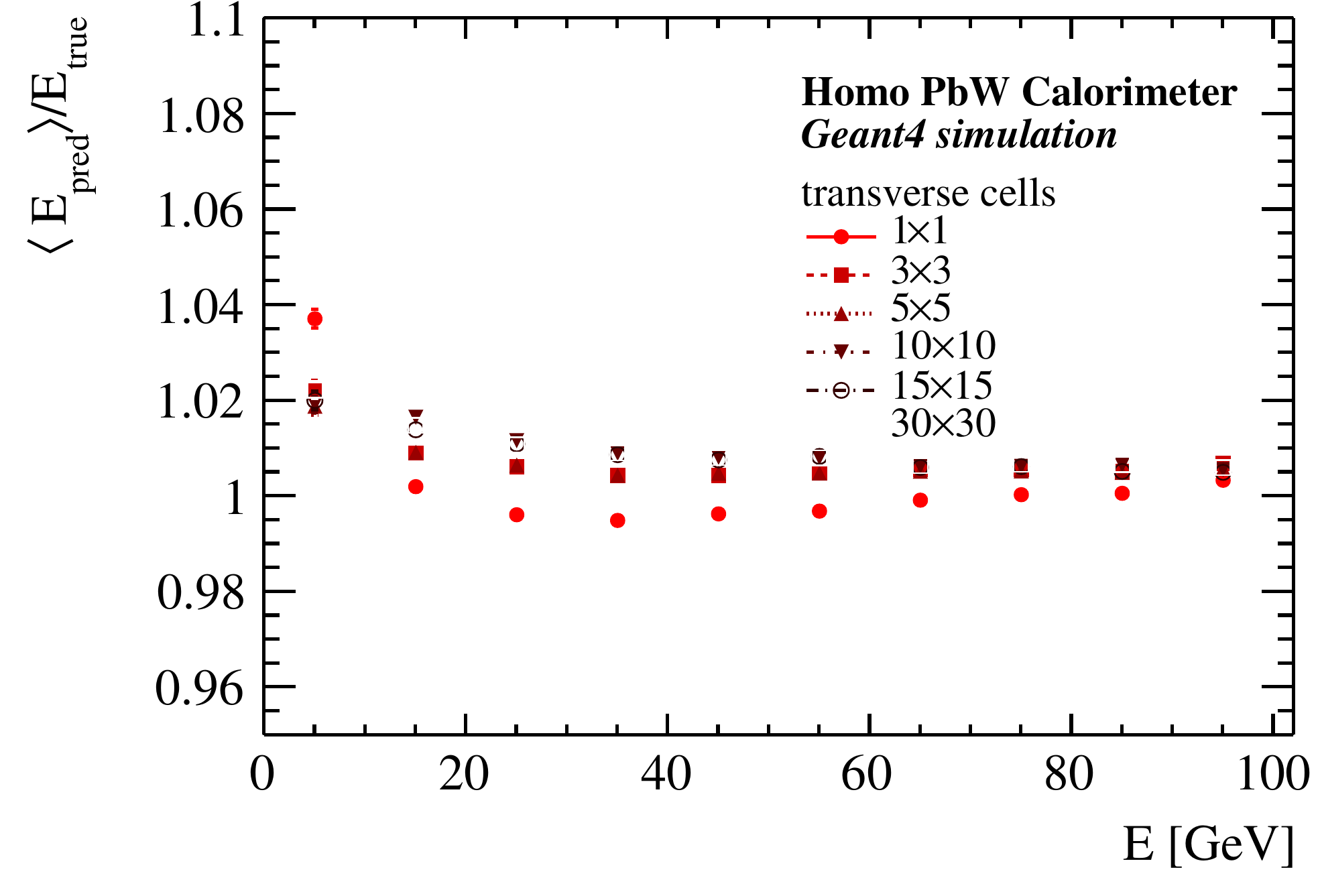}
}
\caption{Energy resolution (a) and linearity (b) for different transverse granularities with 1 longitudinal layer (stage 0). The curves show the fit to the form given in Equation~\ref{eq:reso}.}
\label{fig:trans_0}
\end{center}
\end{figure}

\begin{figure}[htbp]
\begin{center}
\subfloat[]{
\includegraphics[width=0.45\textwidth]{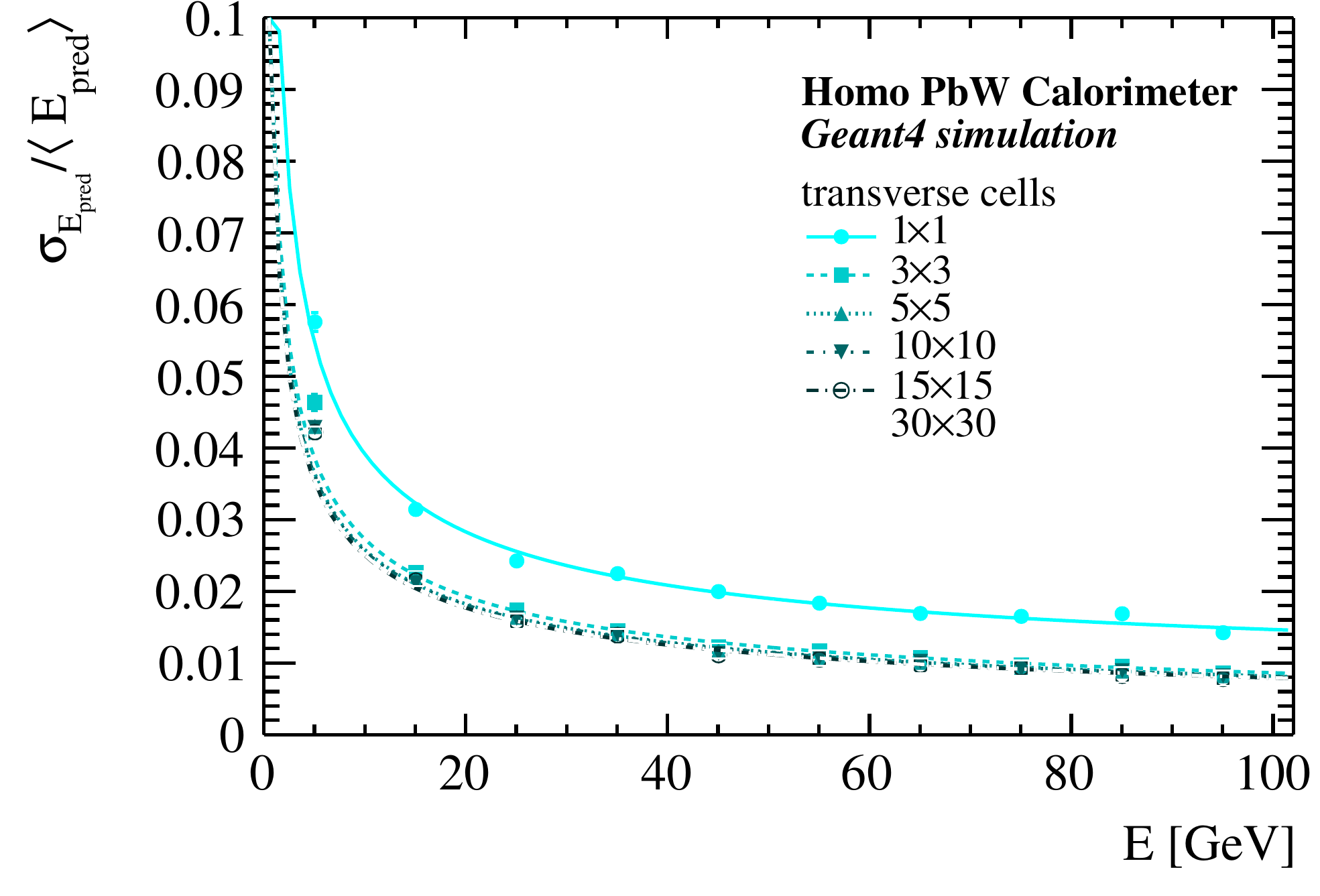}
}\\
\subfloat[]{
\includegraphics[width=0.45\textwidth]{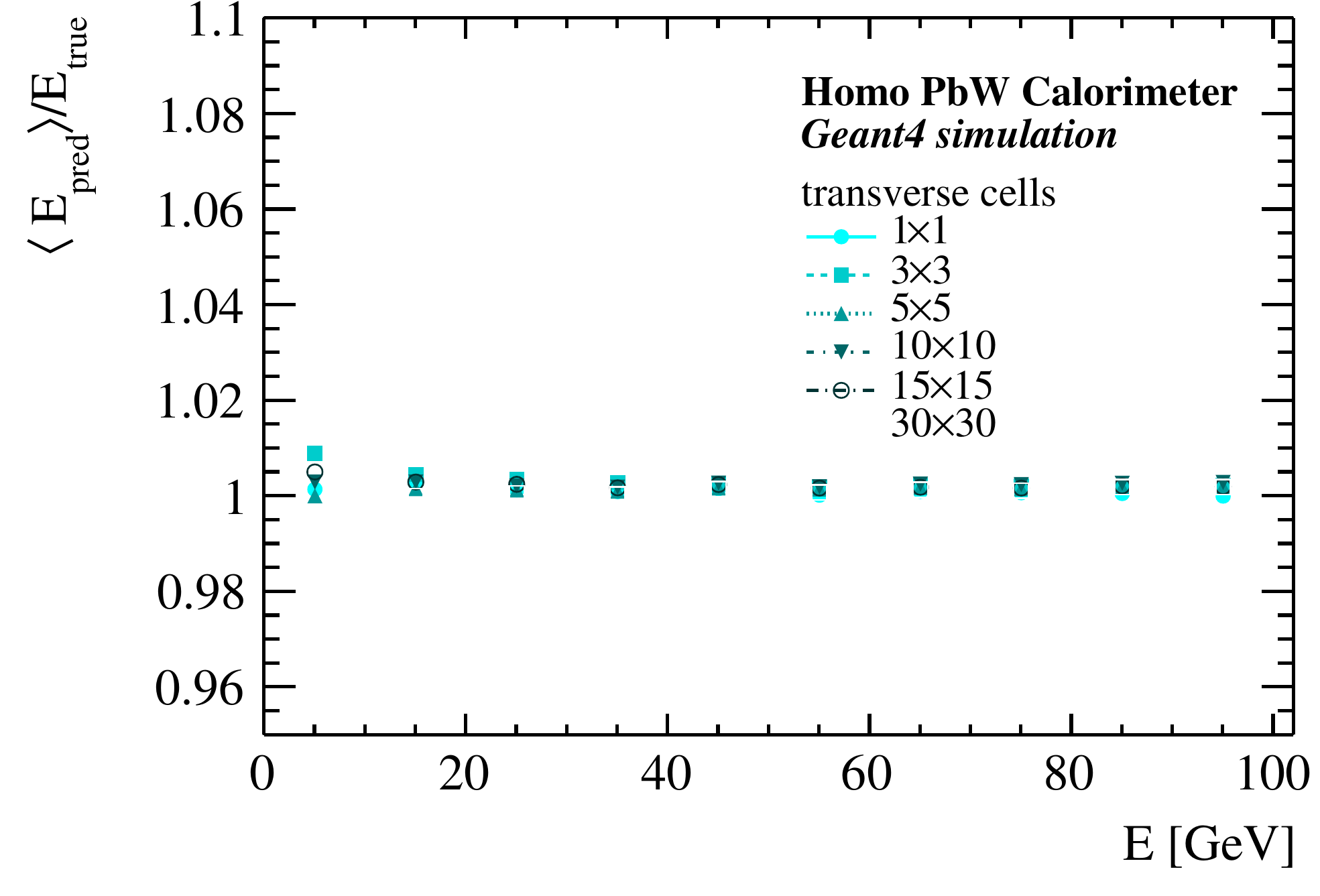}
}
\caption{Energy resolution (a) and linearity (b) for different transverse granularities with 20 longitudinal layers (stage 5). The curves show the fit to the form given in Equation~\ref{eq:reso}.}
\label{fig:trans_5}
\end{center}
\end{figure}

We assume that the DNN is able to identify and re-weight the electromagnetic and hadronic shower fractions, due to the topological differences of EM and hadronic subshowers ($\lambda_{\pi}/X_{0}\sim27$). Thus, we expect the constant term to decrease. Table~\ref{tab:results_reso} shows the resulting measured stochastic and constant terms (using both the Gaussian fit and the RMS to obtain the resolution) for three different sets of scenarios: first, the different longitudinal granularities with no transverse segmentation, the results for which are plotted in Figure~\ref{fig:res}; second, longitudinal stage 0 with different transverse granularities (Figure~\ref{fig:trans_0}); and third, longitudinal stage 5 with different transverse granularities (Figure~\ref{fig:trans_5}). Overall, at the finest granularities, we observe that the constant term goes to zero, while the stochastic term decreases by approximately 50\% with respect to the scenario with no segmentation, reaching a minimum of 8\%, which can be considered the intrinsic stochastic term $\alpha_{\mathrm{int}}$.

\begin{figure}[htbp]
\begin{center}
\subfloat[]{
\includegraphics[width=0.45\textwidth]{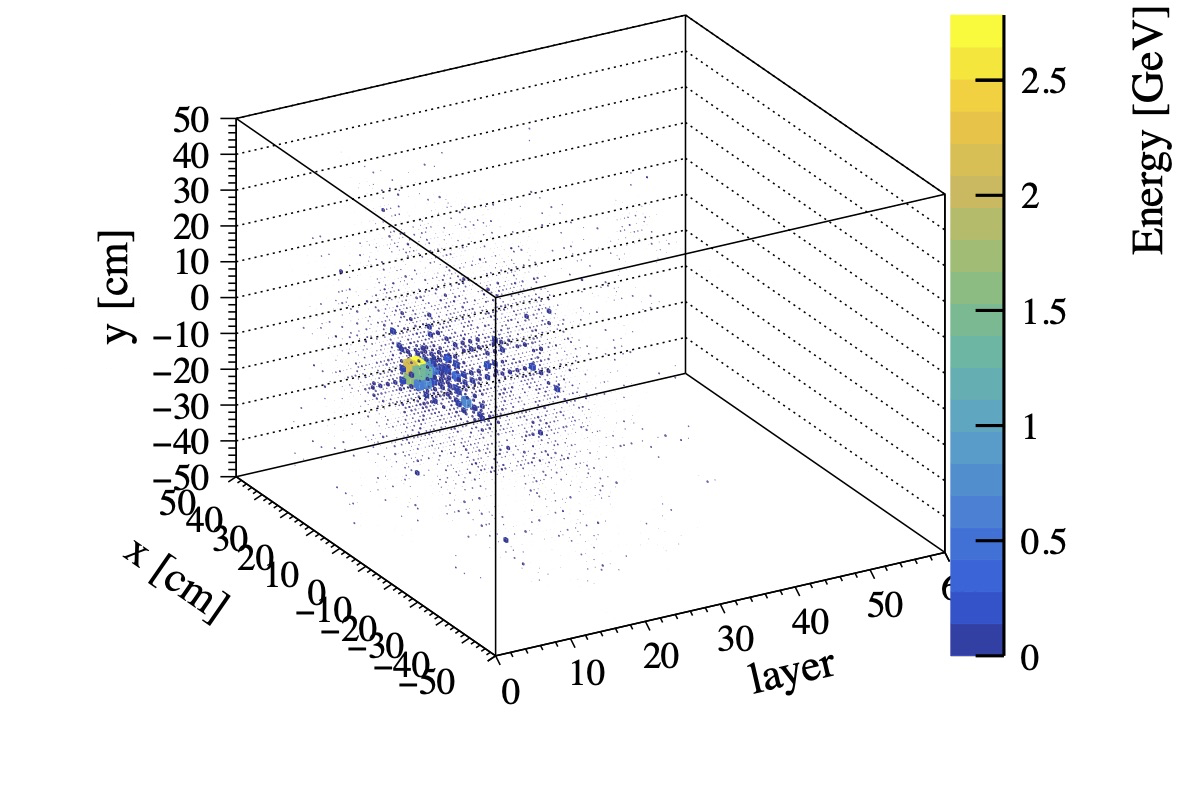}
}\\
\subfloat[]{
\includegraphics[width=0.45\textwidth]{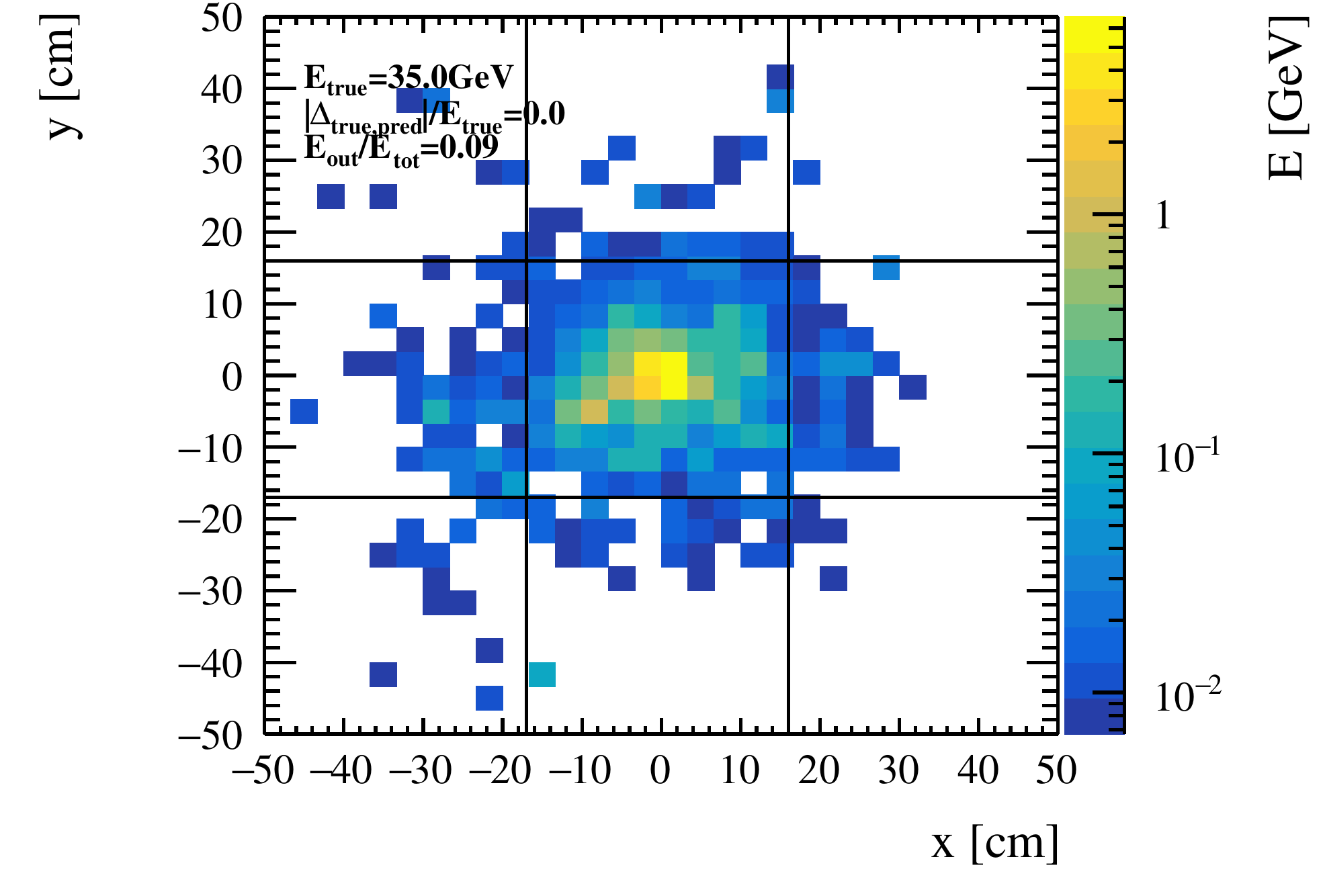}
}
\caption{Event display of a 35\,GeV pion shower in the homogeneous PbW calorimeter. (a) A 3D view at the finest granularity. (b) A front view of the same shower, with a grid overlaid corresponding to the coarsest applied transverse segmentation.}
\label{fig:event}
\end{center}
\end{figure}

The constant term is consistently removed as soon as the first segmentation in transverse granularity into $3\times3$ cells is implemented. Figure~\ref{fig:event} shows an event display of a 35\,GeV pion shower; the bottom shows the impact of a $3\times3$ transverse segmentation. We can see that already at this stage, a significant enough energy fraction of 9\% (shown as $E_{\mathrm{out}}/E_{\mathrm{tot}}$ in the legend) is found in the outer quadrants. In comparison, the same shower is represented in 3D on the top, and visualises the imaging power of the finest chosen granularity of the homogeneous PbW calorimeter.

\begin{table}[htp]
\begin{center}
\begin{tabular}{c|c|c|c|c}
stage		& \multicolumn{2}{c}{stochastic term [\%]}	& \multicolumn{2}{c}{constant term [\%]} \\
			& RMS & Gauss		& RMS & Gauss \\
\hline
0			&20.5 &17.3			&3.0 &2.6			\\
1			&19.7 &15.9			&2.2 &2.0		\\
2			&17.8 &14.0			&1.8 &1.5			\\
3			&17.2 &13.6			&1.7 &1.3			\\
4			&16.0 &12.9			&1.5 &1.1		\\ 
5			&15.4 &12.1			&1.3 &0.8		\\ 
6			&14.6 &11.6			&1.1 &0.6		\\ 
7			&13.0 &10.9			&1.0 &0.5		\\ 
\hline
0A			&20.3 &15.1			&1.3 &0			\\
0B			&20.0 &14.6			&1.2 &0			\\
0C			&18.2 &13.6			&1.3 &0			\\
0D			&18.6 &13.6			&1.1 &0			\\
0E			&17.9 &13.4			&1.3 &0			\\
\hline
5A			&11.4 &8.6			&0.6 &0			\\
5B			&10.6 &8.1			&0.6 &0			\\
5C			&11.0 &8.1			&0 &0			\\
5D			&10.9 &7.9			&0 &0			\\
5E			&10.9 &7.9			&0 &0			\\
\end{tabular}
\end{center}
\caption{Summary of energy resolution fit results. The top set shows the different longitudinal segmentation scenarios with no transverse segmentation, while the other two sets show two specific longitudinal stages with different transverse segmentation scenarios, as described in Table~\ref{tab:dataset}.}
\label{tab:results_reso}
\end{table}%

\begin{figure}[htbp]
\begin{center}
\subfloat[]{
\includegraphics[width=.45\textwidth]{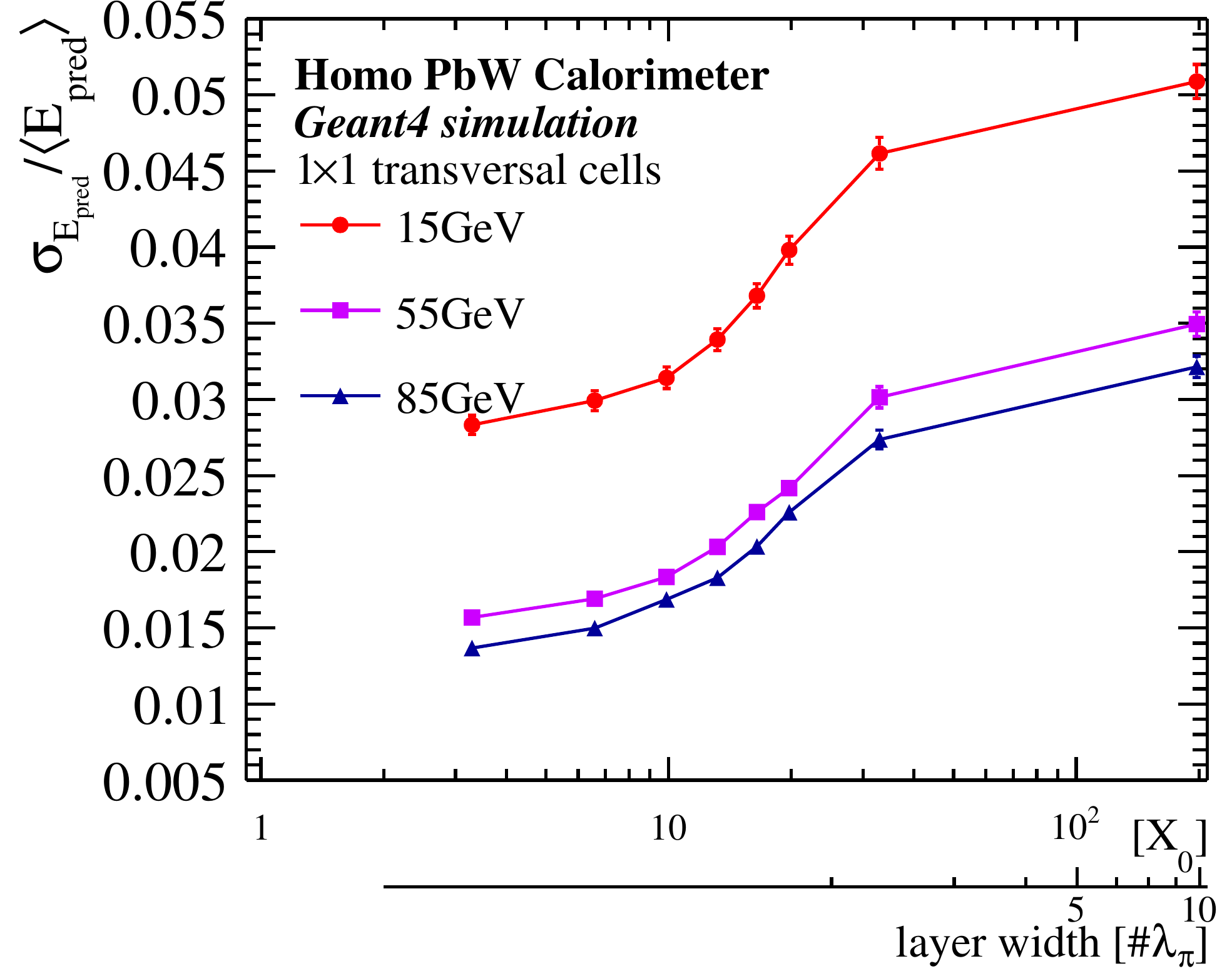}}\\
\subfloat[]{
\includegraphics[width=.45\textwidth]{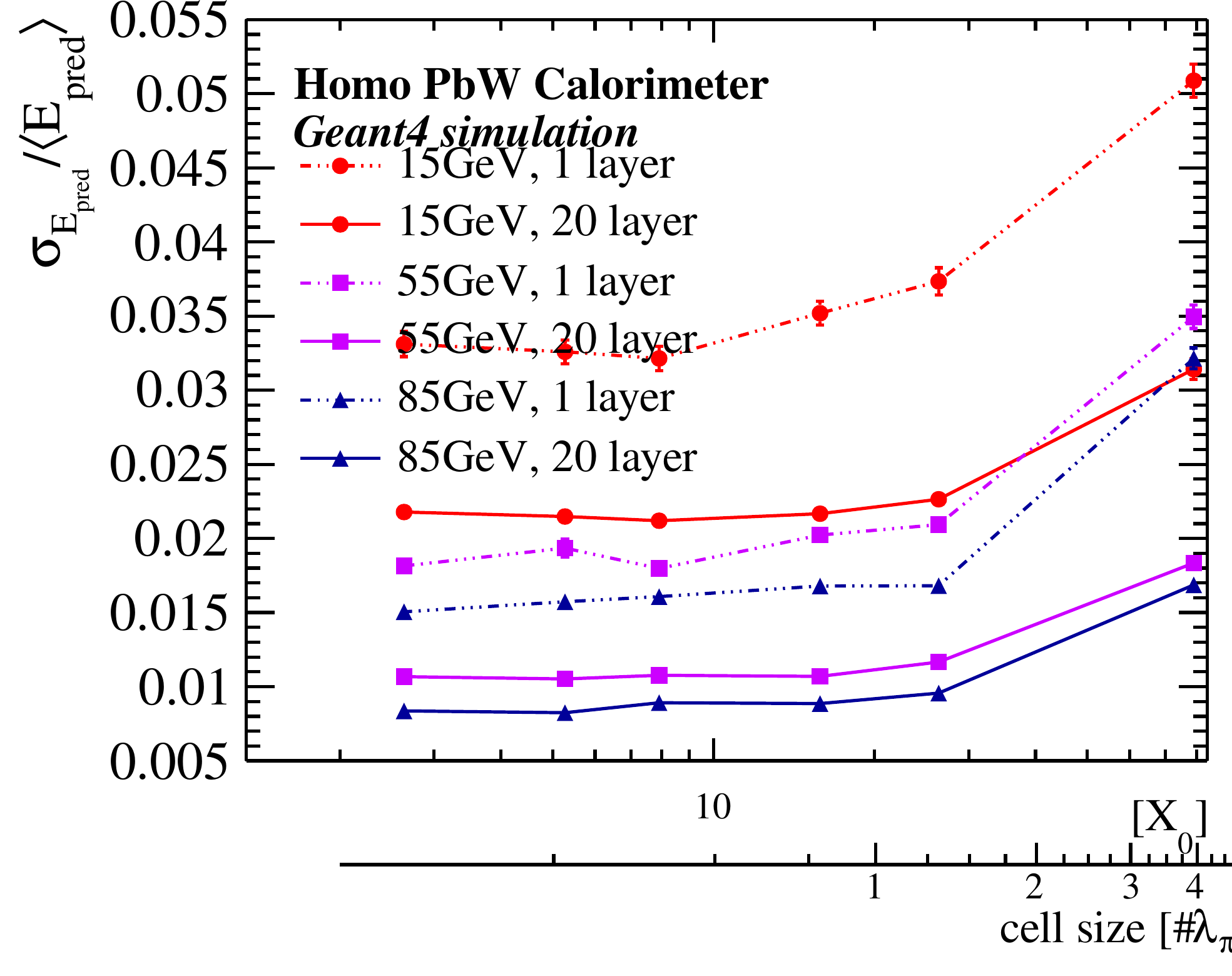}}
\caption{Energy resolution as a function of the longitudinal (a) and transverse (b) granularity. Three different particle energies are considered: 15\,GeV (red circles), 55\,GeV (purple squares), and 85\,GeV (dark blue triangles). In the upper plot, no transverse segmentation is used, while on the bottom, two different longitudinal segmentations are shown: 1 layer (dashed lines) and 20 layers (solid lines).}
\label{fig:summary}
\end{center}
\end{figure}

Figure~\ref{fig:summary} summarizes the energy resolution as a function of longitudinal and transverse granularity. We observe that the behavior of the resolution as a function of granularity exhibits the same pattern regardless of the incident particle energy. For the transverse granularity, the resolution reaches an optimal value at a cell size of $\approx 1\lambda_{\pi}$, and finer segmentation does not yield any appreciable further benefit. In the longitudinal direction, the energy resolution continues to improve as the layer size is decreased, reaching the minimum at the finest granularity considered ($\approx 0.2\lambda_{\pi}$ or $\approx 3 X_0$).

\begin{figure}[htbp]
\begin{center}
\subfloat[]{
\includegraphics[width=.49\textwidth]{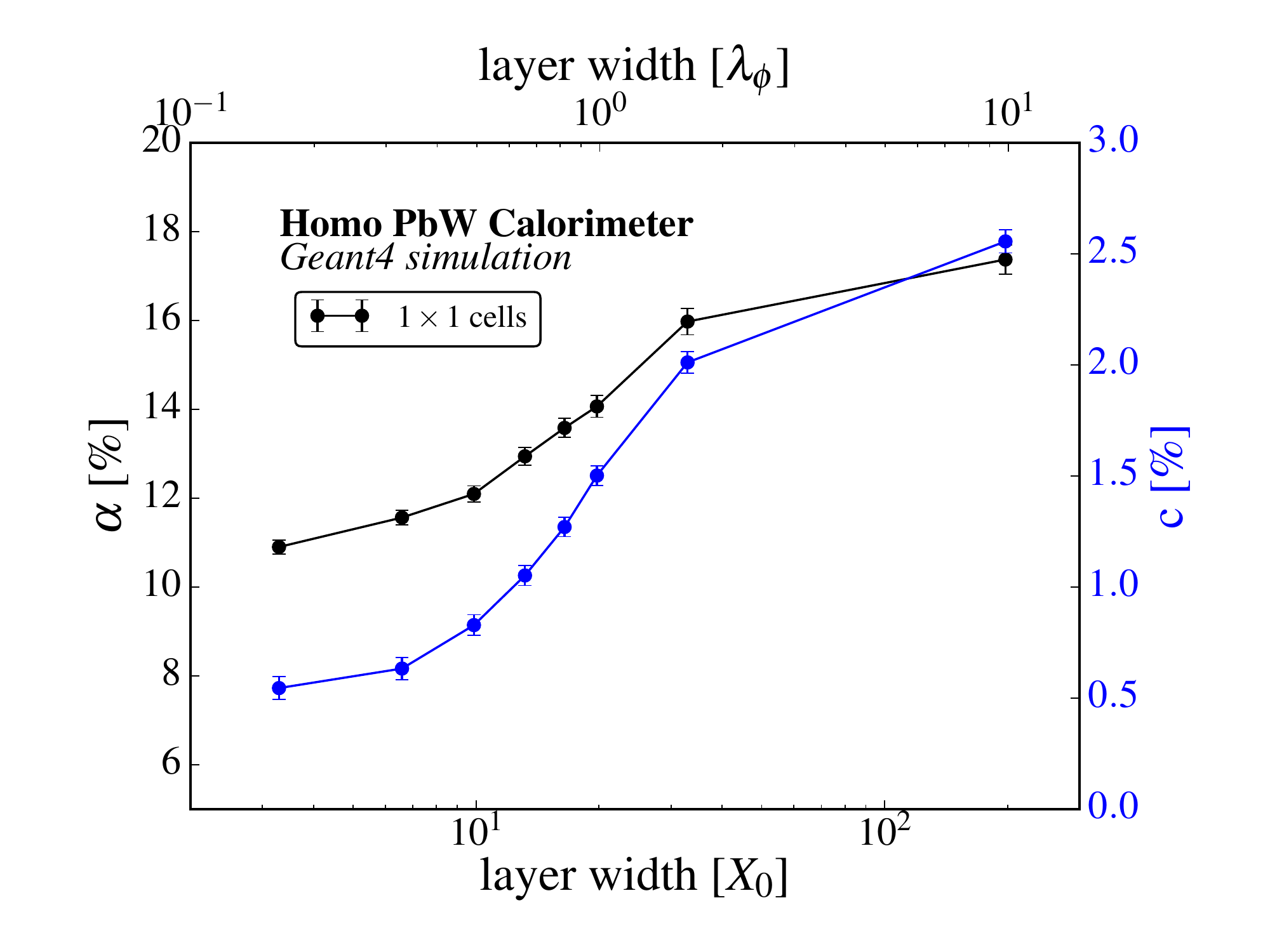}}\\
\subfloat[]{
\includegraphics[width=.49\textwidth]{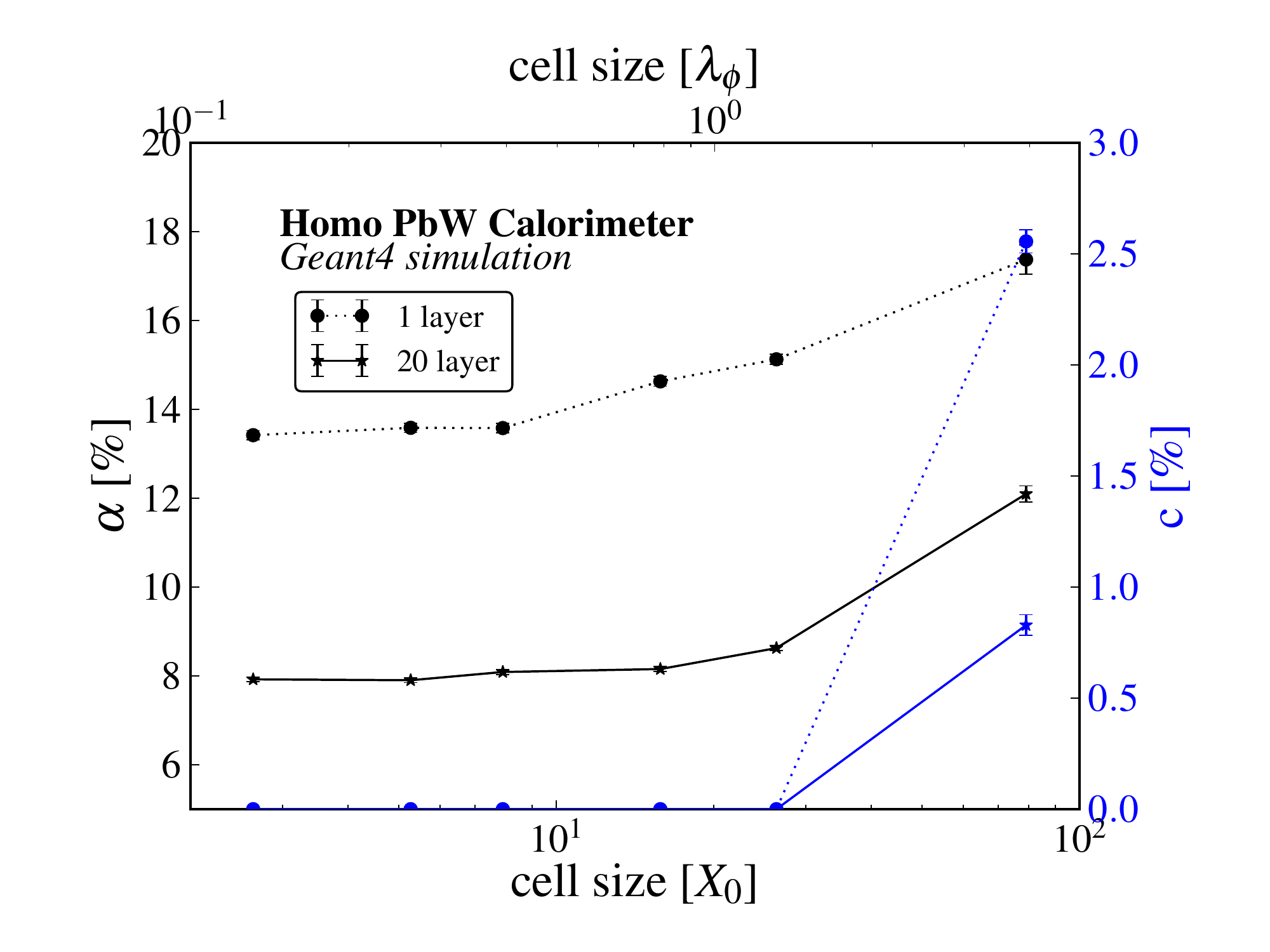}}
\caption{Values of the parameters $\alpha$ (black line) and $c$ (blue line) in the energy resolution function $\frac{\sigma_{E}}{\left<E\right>}=\frac{\alpha}{E} \oplus c$ as a function of the longitudinal (a) and transverse (b) granularity. In the upper plot, no transverse segmentation is used, while on the bottom, two different longitudinal segmentations are shown: 1 layer (dashed lines) and 20 layers (solid lines).}
\label{fig:summary2}
\end{center}
\end{figure}

Figure~\ref{fig:summary2} summarizes the fitted parameters $\alpha$ and $c$ in the energy resolution function in Equation~\ref{eq:reso}, as a function of longitudinal and transverse granularity. In the transverse direction, we observe that the constant term goes to zero at a cell size of $\approx 1 \lambda_{\pi}$ ($25 X_0$), and further decrease in the cell size does not further improve the stochastic term $\alpha$. In the longitudinal case, a layer width in the region 7--10 $X_0$ appears to offer the best balance between the obtained resolution and the detector complexity.

\newpage
\section{Conclusions}
When calorimeters are designed for new high-energy physics experiments, often the approach has been to pick a technology before optimising the reconstruction of jet particles. From the perspective of testing various options, this not only requires significant computing power due to the introduced details of signal processing (digitisation) in the simulations, but also means that the simulations are unable to answer basic questions due to the high complexity. For example, a smaller cell size improves the spatial and pointing resolution, which should help the particle-flow algorithm to reconstruct the jet. However, the signal height per cell decreases, which introduces an energy loss due to a lower signal-to-noise ratio. Thus, a high-level optimisation becomes blind to the individual impact for each effect.
Instead, a different approach could be to first identify the necessary input for reconstruction algorithms which allows for optimal performance, before selecting the detector technology.

Moving towards that approach, we have defined a model calorimeter to identify the necessary cell granularity for a DNN to perform an optimal energy reconstruction. In this model, the impact of the sampling fraction has been intentionally excluded. Even though we are aware that the type of chosen active and passive material will impact the shower development, we believe that this study can be used in order to design a future hadronic calorimeter which allows for optimal energy measurements using DNNs.

These studies suggest that a hadronic calorimeter (with $\lambda_{\pi}/X_{0}\sim27$) should feature cell sizes of at most 1 nuclear interaction length, and longitudinal layers of 7--10\,$X_{0}$ thickness, in order to allow for an optimal software compensation and thus to reach the intrinsic stochastic term of 8\%.
Following this approach, one could imagine further study to determine the optimal cell and layer sizes as a function of the $\lambda_{\pi}/X_{0}$ ratio. However, this exceeds the scope of this paper.

\section*{Acknowledgments}
The training of the models was performed on the GPU clusters of the CERN CMG group. 

\bibliographystyle{spphys}  
\bibliography{references}  

\begin{thebibliography}{10}
\providecommand{\url}[1]{{#1}}
\providecommand{\urlprefix}{URL }
\expandafter\ifx\csname urlstyle\endcsname\relax
  \providecommand{\doi}[1]{DOI \discretionary{}{}{}#1}\else
  \providecommand{\doi}{DOI \discretionary{}{}{}\begingroup
  \urlstyle{rm}\Url}\fi

\bibitem{Sirunyan:2017ulk}
{CMS Collaboration}, JINST \textbf{12}, P10003 (2017).
\newblock \doi{10.1088/1748-0221/12/10/P10003}

\bibitem{Aaboud:2017aca}
{ATLAS Collaboration}, Eur. Phys. J. C \textbf{77}, 466 (2017).
\newblock \doi{10.1140/epjc/s10052-017-5031-2}

\bibitem{Ruan:2014paa}
M.~Ruan, H.~Videau, in \emph{{Proceedings, International Conference on
  Calorimetry for the High Energy Frontier (CHEF 2013): Paris, France, April
  22-25, 2013}} (2013), p. 316

\bibitem{Thomson:2009rp}
M.A. Thomson, Nucl. Instrum. Meth. A \textbf{611}, 25 (2009).
\newblock \doi{10.1016/j.nima.2009.09.009}

\bibitem{Marshall:2012ry}
J.S. Marshall, A.~M{\"u}nnich, M.A. Thomson, Nucl. Instrum. Meth. A
  \textbf{700}, 153 (2013).
\newblock \doi{10.1016/j.nima.2012.10.038}

\bibitem{Marshall:2013bda}
J.S. Marshall, M.A. Thomson, in \emph{{Proceedings, International Conference on
  Calorimetry for the High Energy Frontier (CHEF 2013): Paris, France, April
  22-25, 2013}} (2013), p. 305

\bibitem{Marshall:2015rfa}
J.S. Marshall, M.A. Thomson, Eur. Phys. J. C \textbf{75}, 439 (2015).
\newblock \doi{10.1140/epjc/s10052-015-3659-3}

\bibitem{Sefkow:2015hna}
F.~Sefkow, A.~White, K.~Kawagoe, R.~P{\"o}schl, J.~Repond, Rev. Mod. Phys.
  \textbf{88}, 015003 (2016).
\newblock \doi{10.1103/RevModPhys.88.015003}

\bibitem{Tran:2017tgr}
H.L. Tran, K.~Kr{\"u}ger, F.~Sefkow, S.~Green, J.~Marshall, M.~Thomson,
  F.~Simon, Eur. Phys. J. C \textbf{77}, 698 (2017).
\newblock \doi{10.1140/epjc/s10052-017-5298-3}

\bibitem{CERN-LHCC-97-031}
{CMS Collaboration}, The {CMS} hadron calorimeter project: {Technical Design
  Report}.
\newblock Technical Design Report CERN-LHCC-97-031, {CERN} (1997).
\newblock \urlprefix\url{https://cds.cern.ch/record/357153}

\bibitem{CERN-LHCC-96-041}
{ATLAS Collaboration}, {ATLAS} liquid-argon calorimeter: {Technical Design
  Report}.
\newblock Technical Design Report CERN-LHCC-96-041, {CERN} (1996).
\newblock \urlprefix\url{https://cds.cern.ch/record/331061}

\bibitem{HGCAL-TDR}
{CMS Collaboration}, The {Phase-2} upgrade of the {CMS} endcap calorimeter.
\newblock Technical Design Report CERN-LHCC-2017-023, CMS-TDR-019, {CERN}
  (2017).
\newblock \urlprefix\url{https://cds.cern.ch/record/2293646}

\bibitem{Neubuser:2705432}
C.~Neub{\"u}ser, M.~Aleksa, A.M. Henriques~Correia, J.~Faltova, M.~Selvaggi,
  C.~Helsens, A.~Zaborowska, P.P. Allport, R.R. Bosley, J.~Kieseler,
  A.~Karyukhin, J.S. Schliwinski, N.~Watson, R.R. Stein, A.~Winter,
  O.~Solovyanov, H.F. Pais Da~Silva, J.~Gentil, R.~Goncalo, N.~Topiline,
  Calorimeters for the {FCC-hh}.
\newblock FCC Document CERN-FCC-PHYS-2019-0003, {CERN} (2019).
\newblock \urlprefix\url{https://cds.cern.ch/record/2705432}

\bibitem{Israeli:2018byq}
Y.~Israeli, JINST \textbf{13}(05), C05002 (2018).
\newblock \doi{10.1088/1748-0221/13/05/C05002}

\bibitem{Chefdeville:2019zzq}
{CALICE Collaboration}, Nucl. Instrum. Meth. A \textbf{939}, 89 (2019).
\newblock \doi{10.1016/j.nima.2019.05.013}

\bibitem{Chefdeville:2015}
{CALICE Collaboration}, JINST \textbf{10}, P12006 (2015).
\newblock \doi{10.1088/1748-0221/10/12/P12006}

\bibitem{Adloff:2012gv}
{CALICE Collaboration}, JINST \textbf{7}, P09017 (2012).
\newblock \doi{http://dx.doi.org/10.1088/1748-0221/7/09/P09017}

\bibitem{Eigen:2019ccp}
{CALICE Collaboration}, Nucl. Instrum. Meth. A \textbf{937}, 41 (2019).
\newblock \doi{10.1016/j.nima.2019.04.111}

\bibitem{Quast:2017gnq}
T.~Quast, JINST \textbf{13}(02), C02044 (2018).
\newblock \doi{10.1088/1748-0221/13/02/C02044}

\bibitem{Sefkow:2018rhp}
F.~Sefkow, F.~Simon, J. Phys. Conf. Ser. \textbf{1162}, 012012 (2019).
\newblock \doi{10.1088/1742-6596/1162/1/012012}

\bibitem{CERN-LHCC-97-033}
{CMS Collaboration}, The {CMS} electromagnetic calorimeter project: {Technical
  Design Report}.
\newblock Technical Design Report CERN-LHCC-97-033, {CERN} (1997).
\newblock \urlprefix\url{http://cds.cern.ch/record/349375}

\bibitem{lecun1998gradient}
Y.~LeCun, L.~Bottou, Y.~Bengio, P.~Haffner, Proc. IEEE \textbf{86}, 2278
  (1998).
\newblock \doi{10.1109/5.726791}

\bibitem{ioffe2015batch}
S.~Ioffe, C.~Szegedy, Proc. Mach. Learn. Res. \textbf{37}, 448 (2015).
\newblock \urlprefix\url{http://proceedings.mlr.press/v37/ioffe15.html}

\bibitem{elu_activation}
D.A. Clevert, T.~Unterthiner, S.~Hochreiter, Fast and accurate deep network
  learning by exponential linear units ({ELUs}) (2015).
\newblock {arXiv}:1511.07289

\bibitem{kingma2014adam}
D.P. Kingma, J.~Ba, Adam: A method for stochastic optimization (2014).
\newblock {arXiv}:1412.6980

\bibitem{tensorflow}
M.~Abadi, A.~Agarwal, P.~Barham, E.~Brevdo, Z.~Chen, C.~Citro, et~al.
\newblock {TensorFlow}: Large-scale machine learning on heterogeneous systems.
\newblock \url{https://www.tensorflow.org/} (2015)

\bibitem{keras}
F.~Chollet, et~al.
\newblock Keras.
\newblock \url{https://keras.io} (2015)

\bibitem{DJC}
J.~Kieseler, M.~Stoye, M.~Verzetti, P.~Silva, S.S. Mehta, A.~Stakia, Y.~Iiyama,
  E.~Bols, S.R. Qasim, H.~Kirschenmann, et~al.
\newblock {DeepJetCore} (2020).
\newblock \doi{10.5281/zenodo.3670882}

\bibitem{Fabjan:1989ti}
C.W. Fabjan, R.~Wigmans, Rept. Prog. Phys. \textbf{52}, 1519 (1989).
\newblock \doi{10.1088/0034-4885/52/12/002}

\bibitem{Tiecke:1989nz}
H.~Tiecke, Nucl. Instrum. Meth. A \textbf{277}, 42 (1989).
\newblock \doi{10.1016/0168-9002(89)90533-0}

\end{thebibliography}

%
%
%
%

\end{document}